\documentclass[final]{elsarticle}
\usepackage{a4wide}
\usepackage{amssymb}

\usepackage[utf8]{inputenc}
\usepackage[english]{babel}
\usepackage[T1]{fontenc}
\usepackage{clrscode3e_modified}

\usepackage{amsmath}
\usepackage{color}
\usepackage{multirow}

\usepackage{graphicx,ifpdf}

\usepackage{rotating}
\graphicspath{{figs/}} \ifpdf
\fi

\usepackage[nodots,nocompress]{numcompress}

\usepackage{floatrow}
\usepackage{caption}
\usepackage{subcaption}
\usepackage{calc}
\usepackage{graphicx}

\usepackage{array}

\newtheorem{definition}{Definition}[section]

\biboptions{comma,square}

\journal{Theoretical Computer Science}

\newcolumntype{C}[1]{>{\centering\let\newline\\\arraybackslash\hspace{0pt}}m{#1}}

\newcommand{\changeParkM}[1]{#1}

\newcommand{\changeParkO}[1]{#1}

\begin{document}

\begin{frontmatter}



\title{Order Preserving Matching}


\author[snu]{\mbox{Jinil Kim}}
\ead{jikim@theory.snu.ac.kr}
\author[syu]{\mbox{Peter Eades}}
\ead{peter.eades@sydney.edu.au}
\author[fdu,gut]{\mbox{Rudolf Fleischer}}
\ead{rudolf@fudan.edu.cn}
\author[syu]{\mbox{Seok-Hee Hong}}
\ead{seokhee.hong@sydney.edu.au}
\author[kcl,ctu]{\mbox{Costas S. Iliopoulos}}
\ead{c.iliopoulos@kcl.ac.uk}
\author[snu]{\mbox{Kunsoo Park}\corref{cor1}}
\ead{kpark@theory.snu.ac.kr}
\author[kcl]{\mbox{Simon J. Puglisi}}
\ead{simon.puglisi@kcl.ac.uk}
\author[thu]{\mbox{Takeshi Tokuyama}}
\ead{tokuyama@dais.is.tohoku.ac.jp}

\cortext[cor1]{Corresponding author, Tel:+82-2-880-1828, Fax:+82-2-885-3141}
\address[snu]{Department of Computer Science and Engineering, Seoul National University, South Korea.}
\address[syu]{School of Information Technologies, University of Sydney, Australia.}
\address[fdu]{SCS and IIPL, Fudan University, Shanghai, China.}
\address[gut]{Department of Applied Information Technology, German University of Technology in Oman, Muscat, Oman.}
\address[kcl]{Department of Informatics, King's College London, London, United Kingdom.}
\address[ctu]{Digital Ecosystems and Business Intelligence Institute, Curtin University, Australia.}
\address[thu]{Graduate School of Information Sciences, Tohoku University, Japan.}

\begin{abstract}
We introduce a new string matching problem called \emph{order-preserving matching} on numeric strings where a pattern matches a text if the text contains a substring whose relative orders coincide with those of the pattern. Order-preserving matching is applicable to many scenarios such as stock price analysis and musical melody matching in which the order relations should be matched instead of the strings themselves. Solving order-preserving matching has to do with \emph{representations of order relations} of a numeric string. We define \emph{prefix representation} and \emph{nearest neighbor representation}, which lead to efficient algorithms for order-preserving matching. We present efficient algorithms for single and multiple pattern cases. For the single pattern case, we give an $O(n \log m)$ time algorithm and optimize it further to obtain $O(n + m \log m)$ time. For the multiple pattern case, we give an $O(n \log m)$ time algorithm.
\end{abstract}


\end{frontmatter}


\section{Introduction}

String matching is one of fundamental problems which has been extensively studied in stringology. Sometimes a string consists of numeric values instead of characters in an alphabet, and we are interested in some \emph{trends} in the text rather than specific patterns. For example, in a stock market, analysts may wonder whether there is a period when the share price of a company dropped consecutively for 10 days and then went up for the next 5 days. In such cases, the changing patterns of share prices are more meaningful than the absolute prices themselves. Another example can be found in the melody matching of two musical scores. A musician may be interested in whether her new song has a melody similar to well-known songs. As many variations are possible in a melody where the relative heights of pitches are preserved but the absolute pitches can be changed, it would be reasonable to match relative pitches instead of absolute pitches to find similar phrases.

\begin{figure}[t!]%
\ffigbox
{
    \begin{subfloatrow}%
    \includegraphics[scale=0.62]{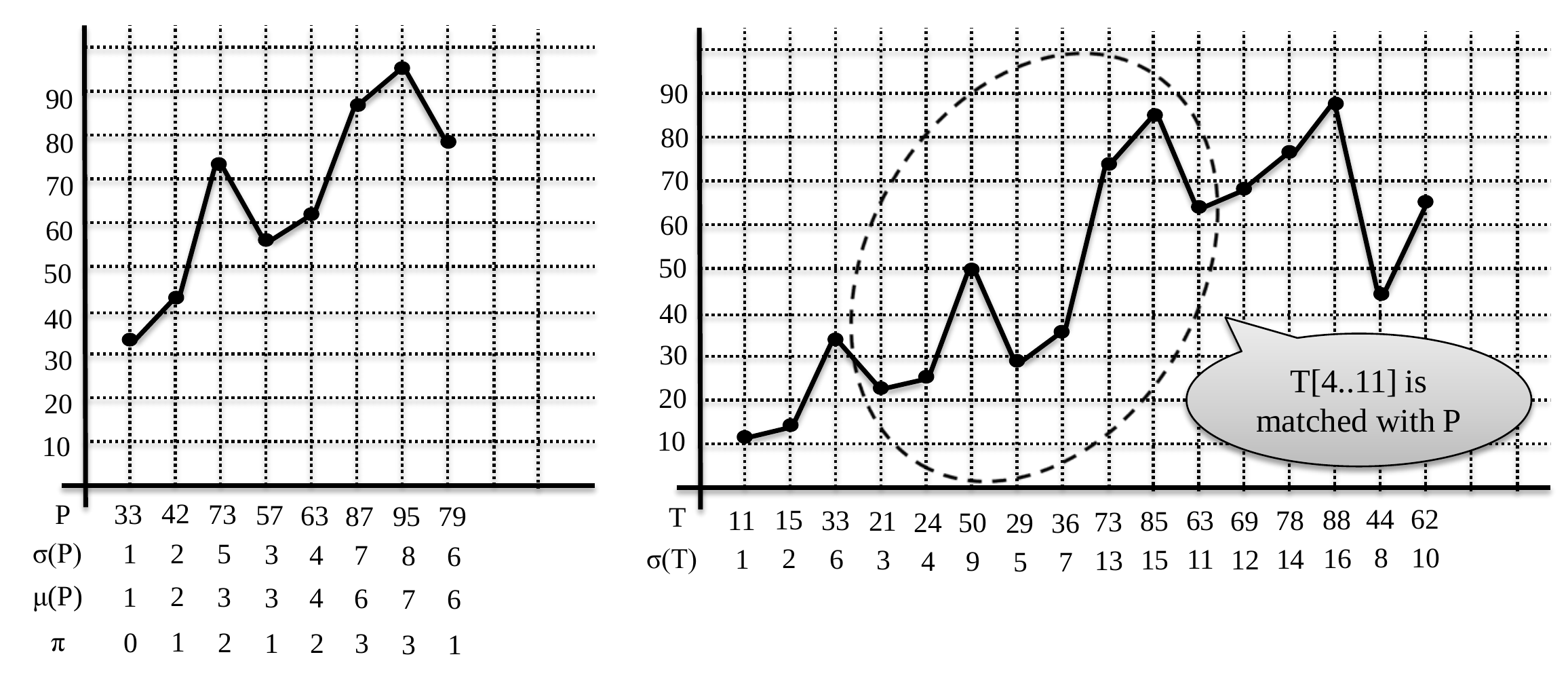}
    \end{subfloatrow}%
}
{\caption{Example of pattern and text}\label{FIG:example_pattern_text}}
\end{figure}

An \emph{order-preserving matching} can be helpful in both examples where a pattern is matched with the text if the text contains a substring whose relative orders coincide with those of the pattern. For example, in Fig.~\ref{FIG:example_pattern_text}, pattern $P=(33,42,73,57,63,87,95,79)$ is matched with text $T$ since the substring $(21,24,50,29,36,73,85,63)$ in the text has the same relative orders as the pattern. In both strings, the first characters $33$ and $21$ are the smallest, the second characters $42$ and $24$ are the second smallest, the third characters $73$ and $50$ are the $5$-th smallest, and so on. If we regard prices of shares or absolute pitches of musical notes as numeric characters of the strings, both examples above can be modeled as order-preserving matching.

Solving order-preserving matching has to do with \emph{representations of order relations} of a numeric string.
If we replace each character in a numeric string by its rank in the string, then we can obtain a (natural) representation of order relations.
But this natural representation is not amenable to developing efficient algorithms because the rank of a character depends on the substring in which the rank is computed.
Hence, we define a \emph{prefix representation} of order relations which leads to an $O(n \log m)$ time algorithm for order-preserving matching where $n$ and $m$ are the lengths of the text and the pattern, respectively.
Surprisingly, however, there is an even better representation, called \emph{nearest neighbor representation}, by which we were able to develop an $O(m \log m + n)$ time algorithm.

In this paper, we define a new class of string matching problem, called \emph{order-preserving matching}, and present efficient algorithms for single and multiple pattern cases. For the single pattern case, we propose an $O(n \log m)$ algorithm based on the Knuth-Morris-Pratt (KMP) algorithm~\cite{CLRS,Cro03}, and optimize it further to obtain $O(n + m \log m)$ time. For the multiple pattern case, we present an $O(n \log m)$ algorithm based on the Aho-Corasick algorithm~\cite{Aho90}.

\textbf{Related Work:} $(\delta, \gamma)$-matching has been studied to search for similar patterns of numeric strings~\cite{CIMP02,CIL+02,CI04,CCI05,LCK06,LMP08,MLP12}. In this paradigm, two parameters $\delta$ and $\gamma$ are given, and two numeric strings of the same length are matched if the maximum difference of the corresponding characters is at most $\delta$ and the total sum of differences is at most $\gamma$. Several variants were studied to allow for \emph{don't care} symbols~\cite{CIL+03}, transposition-invariant~\cite{LCK06} and gaps~\cite{CCF05_WEA,CCF05_ISMIR,FG08}. On the other hand, \changeParkO{some generalized matching problems such as parameterized matching~\cite{Bak93,AFM94}, overlap matching~\cite{ACH+03}, and function matching~\cite{AALP06,AI07} were studied in which \emph{matching} relations are defined differently so that some properties of two strings are matched instead of exact matching of characters~\cite{Mut95}. } However, none of them addresses the \emph{order relations} which we focus on in this paper.


\section{Problem formulation}

\subsection{Notations}

Let $\Sigma$ denote the set of numbers such that a comparison of two numbers can be done in constant time, and let $\Sigma^*$ denote the set of strings over the alphabet $\Sigma$. Let $|x|$ denote the length of a string $x$. A string $x$ is described by either a concatenation of characters $x[1] \cdot x[2] \cdot ... x[|x|]$ or a sequence of characters as $(x[1], x[2], ..., x[|x|])$ interchangeably.
For a string $x$, a substring $x[i..j]$ be $(x[i], x[i+1], ..., x[j])$ and the prefix $x_i$ be $x[1..i]$. The rank of a character $c$ in string $x$ is defined as $rank_x(c) = 1+|\{ i : x[i] < c \; \text{for} \; 1 \leq i \leq |x| \}|$.
For simplicity, we assume that all the numbers in a string are \emph{distinct}. When a number occurs more than once in a string, we can extend our character definition to a pair of character and index in the string so that the characters in the string become distinct.


\subsection{Natural representation of order relations}


For a pattern $P[1..m] \in \Sigma^*$ and a text $T[1..n] \in \Sigma^*$,
a \emph{natural representation} $\sigma(x)$ of the order relations of a string $x$ can be defined as $\sigma(x) = rank_x(x[1]) \cdot rank_x(x[2]) ... \cdot rank_x(x[|x|])$.

\begin{definition}[Order-preserving matching]
Given a text $T[1..n]$ and a pattern $P[1..m]$, $T$ is matched with $P$ at position $i$ if
$\sigma(T[i-m+1..i])=\sigma(P)$.
Order-preserving matching is the problem of finding all positions of $T$ matched with $P$.
\label{definition:order_preserving_matching}
\end{definition}

For example, let's consider two strings $P=(33, 42, 73, 57, 63, 87, 95, 79)$ and $T=(11, 15, 33, 21,$ $24, 50, 29, 36, 73, 85, 63, 69, 78, 88, 44, 62)$ shown in Fig \ref{FIG:example_pattern_text}.
The natural representation of $P$ is $\sigma(P)=(1, 2, 5, 3, 4, 7, 8, 6)$ which is matched with $T[4..11]=(21, 24, 50, 29, 36, 73, 85, 63)$ at position $4$ but is not matched at the other positions of $T$.

As the rank of a character depends on the substring in which the rank is calculated, the string matching algorithms with $O(n+m)$ time complexity such as KMP, Boyer-Moore~\cite{CLRS,Cro03} cannot be applied directly. For example, the rank of $T[4]$ is $3$ in $T[1..8]$ but is changed to $1$ in $T[4..11]$.

The naive pattern matching algorithm is applicable to order-preserving matching if both the pattern and the text are converted to natural representations. If we use \emph{the order-statistic tree} based on the red-black tree~\cite{CLRS}, computing the rank of a character in the string $x$ takes $O(\log |x|)$, which makes the computation time of the natural representation $\sigma(x)$ be $O(|x| \log |x|)$.
The naive order-preserving matching algorithm computes $\sigma(P)$ in $O(m \log m)$ time and
$\sigma(T[i..i+m-1])$ for each position $i \in [1..n-m+1]$ of text $T$ in $O(m \log m)$ time, and compares them in $O(m)$ time. As $n-m+1$ positions are considered, the total time complexity becomes $O((n-m+1) \cdot (m \log m)) = O(nm \log m)$.
As this time complexity is much worse than $O(m+n)$ which we can obtain from the exact pattern matching, sophisticated matching techniques need to be considered for order-preserving matching as discussed in later sections.


\section{$O(n \log m)$ algorithm}

\subsection{Prefix representation}

An alternative way of representing order relations is to use the rank of each character in the prefix. Formally, the \emph{prefix representation} of order relations can be defined as $\mu(x) = rank_{x_1}(x[1]) \cdot rank_{x_2}(x[2]) ... \cdot rank_{x_{|x|}}(x[|x|])$. For example, the prefix representation of $P$ in Fig~\ref{FIG:example_pattern_text} is $\mu(x) = (1,2,3,3,4,6,7,6)$.


An advantage of the prefix representation is that $\mu(x)[i]$ can be computed without looking at characters in $x[i+1..|x|]$ ahead of position $i$. By using the order-statistic tree $\mathcal{T}$ for dynamic order statistics~\cite{CLRS} containing characters of $x[1..i-1]$, $\mu(x)[i]$ can be computed in $O(\log |x|)$ time. Moreover, the prefix representation can be updated incrementally by inserting the next character to $\mathcal{T}$ or deleting the previous character from $\mathcal{T}$. Specifically, when $\mathcal{T}$ contains the characters in $x[1..i]$,
$\mu(x[1..i+1])[i+1]$ can be computed if $x[i+1]$ is inserted to $\mathcal{T}$, and $\mu(x[2..i])[i-1]$ can be computed if $x[1]$ is deleted from $\mathcal{T}$.

Note that there is a \emph{one-to-one} mapping between the natural representation and the prefix representation. The number of all the distinct natural representations for a string of length $n$ is $n!$ which corresponds to the number of permutations, and the number of all the distinct prefix representations is $n!$ too since there are $i$ possible values for the $i$-th character of a prefix representation, which results in $1 \cdot 2 \cdot ... n = n!$ cases.
For any natural representation of a string, there is a conversion function which returns the corresponding prefix representation and vice versa.

The prefix representation of $P$ is easily computable by inserting each character $P[k]$ to $\mathcal{T}$ consecutively as in $\proc{Compute-Prefix-Rep}$. The functions of order-statistic tree are listed up in Fig~\ref{fig:dynamic-order-statistics}. We assume that the index $i$ of $x$ is stored with $x[i]$ in $\proc{OS-Insert}(\mathcal{T}, x, i)$ to support $\proc{OS-Find-Prev-Index}(\mathcal{T}, c)$ and $\proc{OS-Find-Next-Index}(\mathcal{T}, c)$ where the index $i$ of the largest (smallest) character less than (greater than) $c$ is retrieved.

\begin{center}
\small\addtolength{\tabcolsep}{-1pt}
\begin{figure}[t]
  \centering
  \begin{tabular}[ht]{ |l|l| }
\hline Function & Description \\
\hline OS-Insert($\mathcal{T}, x, i$) & Insert $(x[i], i)$ to $\mathcal{T}$\\
\hline OS-Delete($\mathcal{T}, x$) & Delete all the characters of string $x$ from $\mathcal{T}$\\
\hline OS-Rank($\mathcal{T}, c$) & Computes rank $r$ of character $c$ in $\mathcal{T}$\\
\hline OS-Find-Prev-Index($\mathcal{T}, c$) & Find the index $i$ of the largest character less than $c$\\
\hline OS-Find-Next-Index($\mathcal{T}, c$) & Find the index $i$ of the smallest character greater than $c$\\
\hline
\end{tabular}
\caption{List of functions on $\mathcal{T}$ for dynamic order statistics }
\label{fig:dynamic-order-statistics}
\end{figure}
\end{center}

\begin{codebox}
\Procname{$\proc{Compute-Prefix-Rep}(P)$}
\li $m \gets |P|$
\li $\mathcal{T} \gets \phi$
\li $\proc{OS-Insert}(\mathcal{T}, P, 1)$
\li $\mu(P)[1] \gets 1$
\li \For $k \gets 2$ \To $m$
\li \Do
        $\proc{OS-Insert}(\mathcal{T}, P, k)$
\li     $\mu(P)[k] \gets \proc{OS-Rank}(\mathcal{T}, P[k])$
    \End
\li \Return $\mu(P)$
\end{codebox}

The time complexity of $\proc{Compute-Prefix-Rep}$ is $O(m \log m)$ as each of $\proc{OS-Insert}$ and $\proc{OS-Rank}$ takes $O(\log m)$ time and there are $O(m)$ number of such operations.

\subsection{KMP failure function}

The KMP-style failure function $\pi$ of order-preserving matching is well-defined under our prefix representation:

\[
  \pi[q] = \left\{
  \begin{array}{l l}
    \max\{k: \mu(P[1..k])=\mu(P[q-k+1..q]) \; \text{for} \; 1 \le k < q \} & \quad \text{if $q>1$}\\
    0 & \quad \text{if $q=1$}\\
  \end{array} \right.
\]

\begin{figure}[t]
  \includegraphics[scale=0.45]{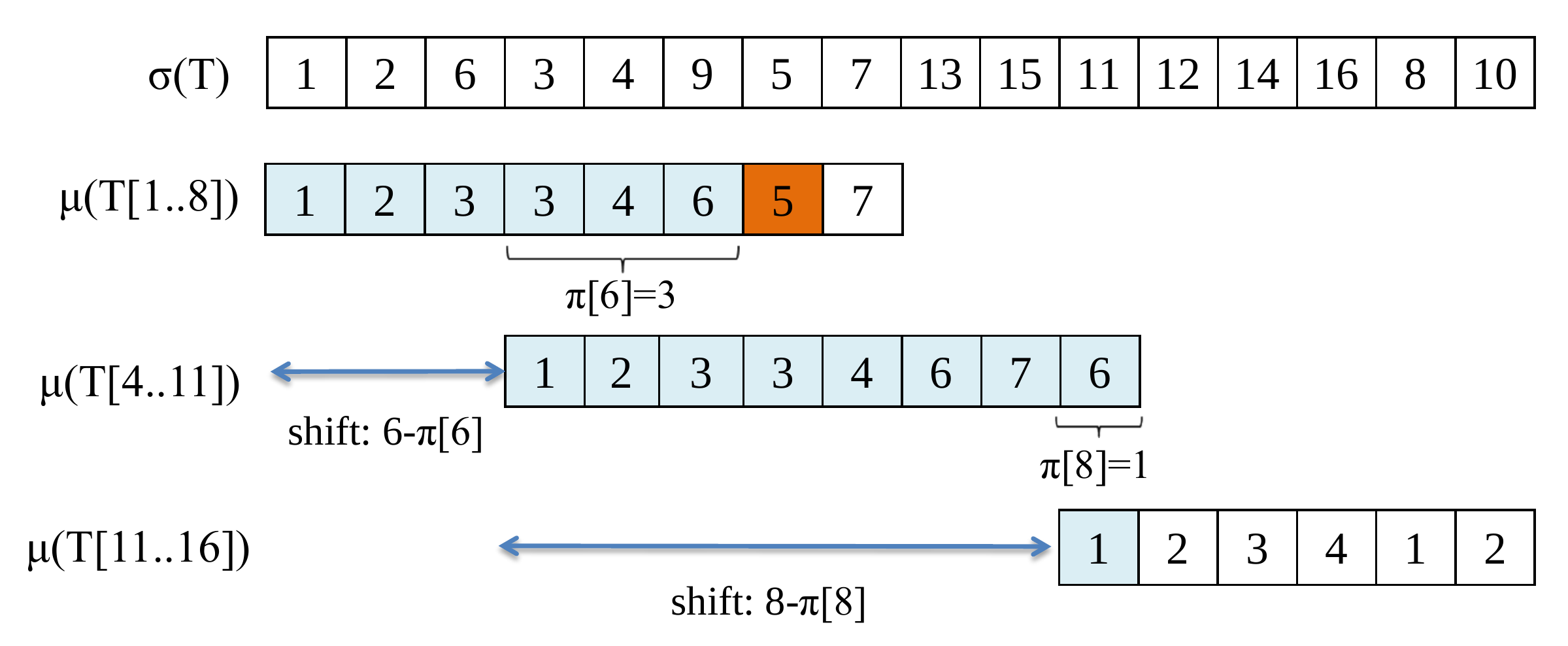}
  \caption{Example of text search}
  \label{FIG:order_matching_example}
\end{figure}

Intuitively, $\pi$ means that the longest proper prefix $\mu(P[1..k])$ of $P$ is matched with $\mu(P[q-k+1..q])$ which is the prefix representation of the suffix of $P[1..q]$ with length $k$.
For example, the failure function of $P$ in Fig~\ref{FIG:example_pattern_text} is $\pi[1..m]=(0, 1, 2, 1, 2, 3, 3, 1)$. As shown in Fig~\ref{FIG:order_matching_example}, $\pi[6]=3$ implies that the longest prefix of $\mu(P[1..8])$ which is matched with the prefix representation of any suffix of $P[1..6]=(33,42,63,57,63,87)$ is $\mu(P[1..\pi[6]])=(1,2,3)$.

The construction algorithm of $\pi$ will be given in section~\ref{construction-of-failure-function}

\subsection{Text search}
\label{text-search}

The failure function $\pi$ can accelerate order-preserving matching by filtering mismatched positions as in the KMP algorithm.
Let's assume that $\mu(P)[1..q]$ is matched with $\mu(T[i-q..i-1])[1..q]$ but a mismatch is found between $\mu(P)[q+1]$ and $\mu(T[i-q..i])[q+1]$.
$\pi[q]$ means that $\mu(P)[1..\pi[q]]$ is already matched with $\mu(T[i-\pi[q]..i-1])[1..\pi[q]]$ and matching can be continued at $P[\pi[q]+1]$ comparing with $T[i]$ from the position $i-\pi[q]$.
Since $P[1..\pi[q]]$ is the longest prefix whose order is matched with the suffix of $T[i-q..i-1]$,
the positions from $i-q$ to $i-\pi[q]-1$ can be skipped without any comparisons as in the KMP algorithm.
Fig~\ref{FIG:order_matching_example} shows how $\pi$ can filter mismatched positions. When $\mu(P)[1..6]$ is matched with $\mu(T[1..6])$
but $\mu(P)[7]$ is different from $\mu(T[1..7])[7]$, we can skip the positions from $1$ to $3$.

$\proc{KMP-Order-Matcher}$ describes the order-preserving matching algorithm assuming that $\mu(P)$ and $\pi$ are efficiently computable. In $\proc{KMP-Order-Matcher}$, for each index $i$ of $T$, $q$ is maintained as the length of the longest prefix of $P$ where $\mu(P)[1..q]$ is matched with $\mu(T)[i-q..i-1]$. At that time, the order-statistic tree $\mathcal{T}$ contains all the characters of $T[i-q..i-1]$. If the rank of $T[i]$ in $\mathcal{T}$ is not matched with that of $P[q+1]$, $q$ is reduced to $\pi[q]$ by deleting all the characters $T[i-q..i-\pi[q]-1]$ from $\mathcal{T}$. If $P[q+1]$ and $T[i]$ have the same rank,
$\mu(P)[1..q+1]=\mu(T)[i-q..i]$ and the length of the matched pattern $q$ is increased by $1$. When $q$ reaches $m$, the relative order of $T[i-m-1..i]$ matches that of $P$.

\begin{codebox}
\Procname{$\proc{KMP-Order-Matcher}(T, P)$}
\li $n \gets |T|$, $m \gets |P|$
\li $\mu(P) \gets \proc{Compute-Prefix-Rep}(P)$
\li $\pi \gets \proc{KMP-Compute-Failure-Function}(P, \mu(P))$
\li $\mathcal{T} \gets \phi$
\li $q \gets 0$
\li \For $i \gets 1$ \To $n$
\li \Do
        $\proc{OS-Insert}(\mathcal{T}, T, i)$
\li     $r \gets \proc{OS-Rank}(\mathcal{T}, T[i])$
\li     \While $q > 0$ and $r \neq \mu(P)[q+1]$
                                                                \label{li:KMP-Order-Matcher-while}
\li     \Do
            $\proc{OS-Delete}(\mathcal{T}, T[i-q..i-\pi[q]-1])$
\li         $q \gets \pi[q]$
\li         $r \gets \proc{OS-Rank}(\mathcal{T}, T[i])$
        \End
\li     $q \gets q+1$
                                                                \label{li:KMP-Order-Matcher-inc_q}
\li     \If $q \isequal m$
\li     \Then
            print ``pattern occurs at position" $i$
\li         $q \gets \pi[q]$
        \End
    \End
\end{codebox}

$\proc{KMP-Order-Matcher}$ is different from the KMP algorithm of the exact pattern matching in that the matches are done by order relations instead of characters.
For each position $i$ of $T$, the prefix representation $\mu(T[i-q..i])[q+1]$ of $T[i]$ is computed using order-statistic tree $\mathcal{T}$. If $\mu(T[i-q..i])[q+1]$ does not match $\mu(P)[q+1]$, $q$ is reduced to $\pi[q]$ so that $P$ implicitly shifts right by $q-\pi[q]$.

Another subtle difference is that we do not check whether $r \isequal \mu(P)[q+1]$ before increasing $q$ by $1$ in line~\ref{li:KMP-Order-Matcher-inc_q} (cf.~\cite{CLRS,Cro03}) because it should be satisfied automatically.
From the condition of the while loop in line~\ref{li:KMP-Order-Matcher-while}, $q=0$ or $r = \mu(P)[q+1]$ in line~\ref{li:KMP-Order-Matcher-inc_q}, and if $q=0$, $\mu(P)[1]=1$ for any pattern and it matches any text of length $1$.

The time required in $\proc{KMP-Order-Matcher}$ except the computation of the prefix representation of $P$ and the construction of the failure function $\pi$
can be analyzed as follows. Each $\proc{OS-Insert}$, $\proc{OS-Rank}$ can be done in $O(\log m)$ time while $\proc{OS-Delete}$ in $O(\log m)$ time per deleting each character.
The number of $\proc{OS-Insert}$ is $n$, and the number of deletions is at most $n$, which makes the total time of deletions $O(n \log m)$.
In the same way, the number of $\proc{OS-Rank}$ is bounded by $2n$, $n$ for new characters, and the other $n$ for the computation of rank after reducing $q$, the total cost of $\proc{OS-Rank}$ is also $O(n \log m)$. To sum up, the time for $\proc{KMP-Order-Matcher}$ can be bounded by $O(n \log m)$ except the external functions.

\subsection{Construction of KMP failure function}
\label{construction-of-failure-function}

The construction of failure function $\pi$ can be done similarly to the text match as in the KMP algorithm
where each element $\pi[q]$ is computed by using the previous values $\pi[1..q-1]$.

$\proc{KMP-Compute-Failure-Function}$ describes the construction algorithm of $\pi$. It first tries to
compute $\pi[q]$ starting from the match of $\mu(P[1..\pi[q-1]])$ and $\mu(P[q-\pi[q-1]..q-1])$.
If $\mu(P[1..\pi[q-1]+1)[\pi[q-1]+1] = \mu(P[q-\pi[q-1]..q])[\pi[q-1]+1]$, set $\pi[q]=\pi[q-1]+1$.
Otherwise, it tries another match for $\pi[\pi[1..q-1]]$, and repeats until $\pi[q]$ is computed.

\begin{figure}[t]
  \includegraphics[scale=0.45]{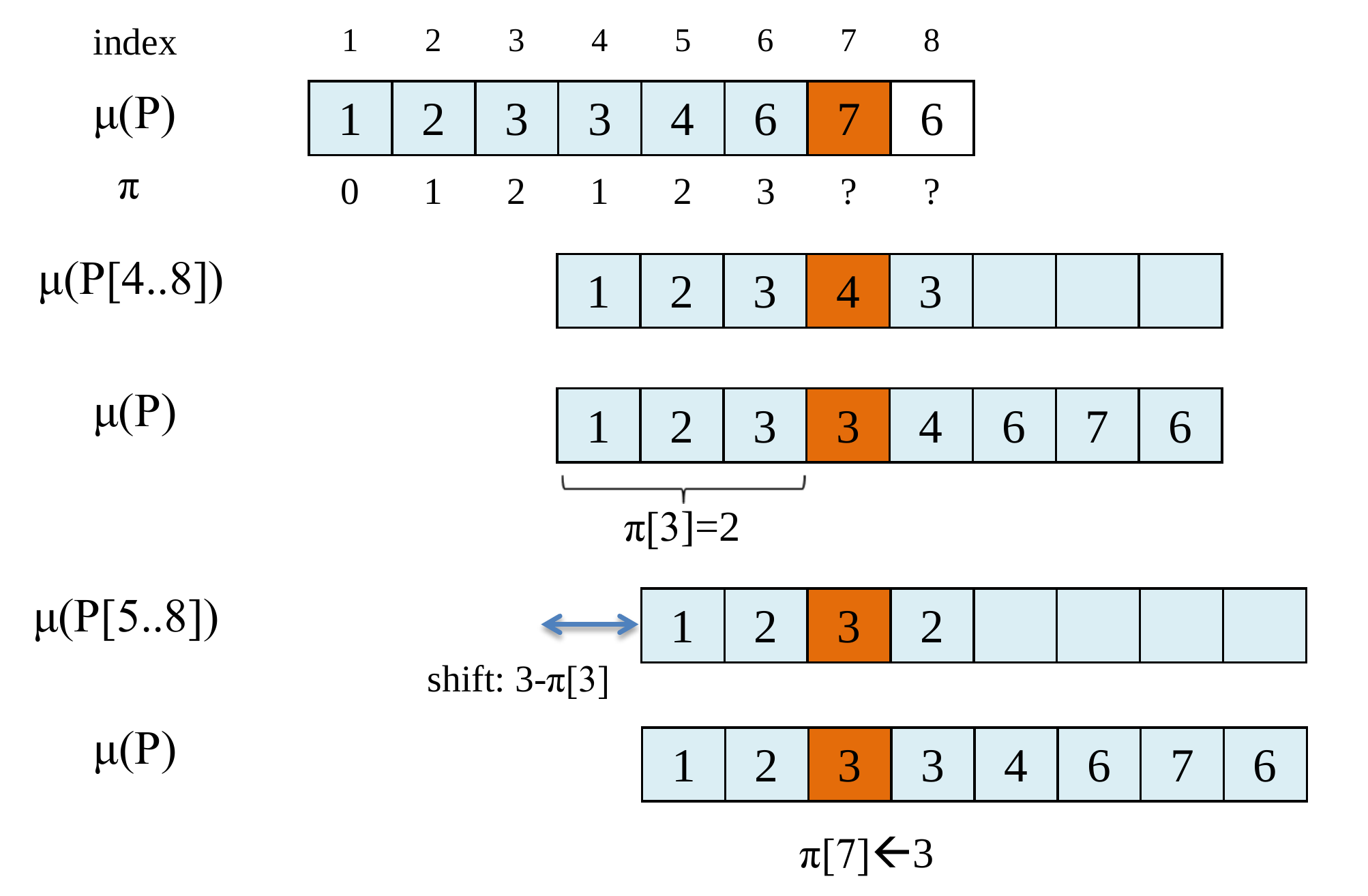}
  \caption{Example of computing failure function}
  \label{FIG:order_failure_function_example_resize}
\end{figure}

Fig~\ref{FIG:order_failure_function_example_resize} shows an example of computing failure functions on $P$ in Fig.~\ref{FIG:example_pattern_text} in which $\pi[7]$ is being computed. Starting from $q=\pi[6]=3$, $\proc{KMP-Order-Matcher}$ tries to match $\mu(P[4..8])[4]$ with $\mu(P)[4]$ but it fails. Then, $q$ is decreased to $q=\pi[3]=2$ and it tries to match $\mu(P[5..8])[3]$ with $\mu(P)[3]$  and it succeeds. $\pi[7]$ is assigned to $\pi[3]+1$, and the next iteration is started with $q=\pi[7]$.

\begin{codebox}
\Procname{$\proc{KMP-Compute-Failure-Function}(P, \mu(P))$}
\li $m \gets |P|$
\li $\mathcal{T} \gets \phi$
\li $\proc{OS-Insert}(\mathcal{T}, P, 1)$
\li $k \gets 0$
\li $\pi[1] \gets 0$
\li \For $q \gets 2$ \To $m$
\li \Do
        $\proc{OS-Insert}(\mathcal{T}, P, q)$
\li     $r \gets \proc{OS-Rank}(\mathcal{T}, P[q]$)
\li     \While $k > 0$ and $r \neq \mu(P)[k+1]$
\li     \Do
            $\proc{OS-Delete}(\mathcal{T}, P[q-k..q-\pi[k]-1])$
\li         $k \gets \pi[k]$
\li         $r \gets \proc{OS-Rank}(\mathcal{T}, P[q]$)
        \End
\li     $k \gets k+1$
\li     $\pi[q] \gets k$
        \End
\li     \Return $\pi$
\end{codebox}

%
%
%

The time complexity of $\proc{KMP-Compute-Failure-Function}$ can be analyzed as in that of $\proc{KMP-Order-Matcher}$
by replacing the length of $T$ with the length of $P$, which results in $O(m \log m)$ time.

\subsection{Correctness and time complexity}

The correctness of our matching algorithm comes from the fact that the failure function is well defined as in the KMP algorithm.
From the analysis of Section~\ref{text-search} and \ref{construction-of-failure-function}, it is clear that our algorithm does not miss any matching position.


The total time complexity is $O(n \log m)$ due to $O(m \log m)$ for prefix representation and failure function computation, $O(n \log m)$ for text search.
Compared with $O(n)$ time of the exact pattern matching, our algorithm has the overhead of $O(\log m)$ factor, which can be optimized at the subsequent section.

\subsection{\changeParkO{Remark on the Boyer-Moore approach}}

Variants of the Boyer-Moore algorithm~\cite{BM77,Hor80,Sun90} may be designed for order-preserving matching in which case the prefix representation should be replaced by the \emph{suffix} representation to proceed matching from right to left of the pattern. The good suffix heuristic~\cite{BM77} is well-defined with the suffix representation, but the bad character heuristic~\cite{BM77} is not applicable since
the character itself has nothing to do with order relations.
As the performance of the Boyer-Moore algorithm is significantly dependent on the bad character heuristic, we cannot expect that the gain of Boyer-Moore variants for order-preserving matching is comparable to that of the original Boyer-Moore algorithm for the exact matching.
Moreover, some practical algorithms such as the Horspool~\cite{Hor80} and the Sunday algorithms~\cite{Sun90} cannot be applied to order-preserving matching because they employ only the bad character heuristic for filtering mismatched positions.

\section{$O(n + m \log m)$ algorithm}

\subsection{Nearest neighbor representation}

The text search of the previous algorithm can be optimized further to remove $O(\log m)$ overhead of computing rank functions.
In the text search of the $O(n \log m)$ algorithm, the rank of each character $T[i]$ in $T[i-q-1..i]$ is computed to check whether it is matched with
$\mu(P)[q+1]$ when we know that $\mu(P)[1..q]$ is matched with $\mu(T[i-q+1..i])$. If we can do it directly without computing
$\mu(P)[q+1]$, the overhead of the operations on $\mathcal{T}$ can be removed.

The main idea is to check whether the order of each character in the text matches that of the corresponding character in the pattern by comparing characters themselves without computing rank values explicitly. When we need to check if a character $x[i]$ of string $x$ has a specific rank value $r$ in prefix $x_i$, we can do it by checking $x[j]<x[i]<x[k]$ where $x[j]$ and $x[k]$ are characters with the nearest rank values of $r$.


The \emph{nearest neighbor representation} of the order relations can be defined as follows. For string $x$, $\nu_p(x)[1..|x|]$ and $\nu_n(x)[1..|x|]$ are the nearest neighbor representation of $x$ where $\nu_p(x)[i]$ is the index of the largest character of $x_{i-1}$ less than $x[i]$ and $\nu_n(x)[i]$ is the index of the smallest character of $x_{i-1}$ greater than $x[i]$. Let $\nu_p(x)[i]=-\infty$ if there is no character less than $x[i]$ in $x_{i-1}$ and let $\nu_n(x)[i]=\infty$ if there is no character greater than $x[i]$ in $x_{i-1}$. Let $x[-\infty]=-\infty$ and $x[\infty]=\infty$.

\begin{center}
\small\addtolength{\tabcolsep}{-1pt}
\begin{figure}[t]
  \centering
\begin{tabular}[ht]{ c C{0.7cm} C{0.7cm} C{0.7cm} C{0.7cm} C{0.7cm} C{0.7cm} C{0.7cm} C{0.7cm}  }
\hline $i$           & $1$       & $2$       & $3$ & $4$ & $5$ & $6$ & $7$ & $8$\\
\hline\hline $\mu(x)[i]$   & $1$       & $2$       & $3$ & $3$ & $4$ & $6$ & $7$ & $6$\\
\hline $\pi[i]$      & $0$       & $1$       & $2$ & $1$ & $2$ & $3$ & $3$ & $1$\\
\hline $\nu_p(x)[i]$ & $-\infty$ & $1$       & $2$ & $2$ & $4$ & $3$ & $6$ & $3$\\
\hline $\nu_n(x)[i]$ & $\infty$ & $\infty$ & $\infty$ & $3$ & $3$ & $\infty$ & $\infty$ & $6$\\
\hline
\end{tabular}
\caption{Example of the nearest neighbor representation}
\label{fig:nearest-neighbor-representation}
\end{figure}
\end{center}

The advantage of the nearest neighbor representation is that we can check whether each text character is matched with the corresponding pattern character in constant time without computing rank explicitly. Fig \ref{fig:nearest-neighbor-representation} shows the nearest neighbor representation of the order relations of $P$ in Fig \ref{FIG:example_pattern_text}. Suppose that $\mu(P)[1..i-1] = \mu(T[1..i-1])$ for $1 \leq i \leq m$. If $T[\nu_p(x)[i]] < T[i] < T[\nu_n(x)[i]]$, then $\mu(P[1..i]) = \mu(T[1..i])$. For example, $\mu(T[1])[1]$ must be matched with $\mu(P)[1]$ since $T[\nu_p(x)[1]] < c < T[\nu_n(x)[1]]$ for any character $c$, which coincides with the fact that the rank in the text of size $1$ is always $1$.
For the second character, $\mu(x)[2]=2$ and $T[2]$ should be larger than $T[1]$ to have $\mu(T[1..2])[2]=2$, which is represented by
$\nu_p(x)[1]=1$ and $\nu_n(x)[1]=\infty$. In this way, for each character, we can decide whether the order of $T[i]$ in $\mu(T[1..i])$ is matched with that of $P[i]$ in $\mu(P[1..i])$ by checking $T[\nu_p(x)[i]] < T[i] < T[\nu_n(x)[i]]$.

$\proc{Compute-Nearest-Neighbor-Rep}$ describes the construction of the nearest neighbor representation of the string $P$
where $\mathcal{T}$ contains the characters of $P_{k-1}$ in each step of the loop.
We assume that $\proc{OS-Find-Prev-Index}(\mathcal{T}, c)$(and
$\proc{OS-Find-Next-Index}(\mathcal{T}, c)$) returns the index $i$ of the largest (smallest) character less than (greater than) $c$, and returns $-\infty$ ($\infty$) if there is no such character.

\begin{codebox}
\Procname{$\proc{Compute-Nearest-Neighbor-Rep}(P)$}
\li $m \gets |P|$
\li $\mathcal{T} \gets \phi$
\li $\proc{OS-Insert}(\mathcal{T}, P, 1)$
\li $(\nu_p(P)[1], \nu_n(P)[1]) \gets (-\infty, \infty)$
\li \For $k \gets 2$ \To $m$
\li \Do
        $\proc{OS-Insert}(\mathcal{T}, P, k)$
\li     $\nu_p(P)[k] \gets \proc{OS-Find-Prev-Index}(\mathcal{T}, P[k])$
\li     $\nu_n(P)[k] \gets \proc{OS-Find-Next-Index}(\mathcal{T}, P[k])$
    \End
\li \Return $(\nu_p(P), \nu_n(P))$
\end{codebox}


The time complexity of $\proc{Compute-Nearest-Neighbor-Rep}$ is $O(m \log m)$ since it has $m$ iterations of the loop and there are $3$ function calls on the order-statistic tree $\mathcal{T}$ taking $O(\log m)$ time in each iteration.

\subsection{Text search}

With the nearest neighbor representation of pattern $P$ and the failure function $\pi$, we can simplify text search so that it does not employ $\mathcal{T}$ at all. For each character $T[i]$, we can check $\mu(P)[q+1]=\mu(T[i-q..i])[q+1]$ by comparing $T[i]$ with
the characters in $T[i-q..i]$ whose indexes correspond to $\nu_p(P)[q+1]$ and $\nu_n(P)[q+1]$ in $P$.
Specifically, if $T[i-q+\nu_p(P)[q+1]-1] < T[i] < T[i-q+\nu_n(P)[q+1]-1]$, then $\mu(P)[q+1]=\mu(T[i-q..i])[q+1]$ must be satisfied
since the relative order of $T[i]$ in $T[i-q..i]$ is the same with that of $P[q+1]$ in $P[1..q+1]$.

For example, let's come back to the text matching example in Fig~\ref{FIG:order_matching_example}. When $\mu(P)[1..6]$ is matched with $\mu(T[1..6])$, we can check $\mu(T[1..7])[7]$ is matched with $\mu(P)[7]$ by checking if $T[7-6+\nu_p(P)[7]-1] < T[7] < T[7-6+\nu_n(P)[7]-1]$, which can be done in constant time. As $T[6]=50$, $T[\infty]=\infty$ but $T[7]=29$, $T[7]$ should have a rank lower than $\mu(P)[7]$, thus $\mu(T[1..7])$ cannot be matched with $\mu(P)[1..7]$.

$\proc{KMP-Order-Matcher2}$ describes the text search algorithm using the nearest neighbor representation.
The algorithm is essentially equivalent to the previous one but simpler since no rank function has to be calculated explicitly.

\begin{codebox}
\Procname{$\proc{KMP-Order-Matcher2}(T, P)$}
\li $n \gets |T|$, $m \gets |P|$
\li $(\nu_{p}(P), \nu_{n}(P)) \gets \proc{Compute-Nearest-Neighbor-Rep}(P)$
\li $\pi \gets \proc{KMP-Compute-Failure-Function2}(P, \nu_{p}(P), \nu_{n}(P))$
\li $q \gets 0$
\li \For $i \gets 1$ \To $n$
\li \Do
        $(j_1,j_2) \gets (\nu_{p}(P)[q+1], \nu_{n}(P)[q+1])$
\li     \While $q > 0$ and ($T[i]<T[i-q+j_1-1]$ or $T[i]>T[i-q+j_2-1])$)
\li     \Do
            $q \gets \pi[q]$
\li         $(j_1,j_2) \gets (\nu_{p}(P)[q+1], \nu_{n}(P)[q+1])$
        \End
\li     $q \gets q+1$
\li     \If $q \isequal m$
\li     \Then
            print ``pattern occurs at position" $i$
\li         $q \gets \pi[q]$
        \End
    \End
\end{codebox}

%
%

The time complexity of $\proc{KMP-Order-Matcher2}$ except the precomputation of the prefix representation and the failure function is $O(n)$ because only one scan of the text is required in the for loop as in the KMP algorithm.

\subsection{Construction of KMP failure function}
\label{NN-construction-of-failure-function}

The construction of the failure function $\pi$ is an extension of $\proc{KMP-Compute-Failure-Function}$ in section~\ref{construction-of-failure-function}
where the rank functions on $\mathcal{T}$ is replaced with comparison of characters using $\nu_{p}(P)$ and $\nu_{n}(P)$
as in $\proc{KMP-Order-Matcher2}$.
$\proc{KMP-Compute-Failure-Function2}$ describes the construction of the KMP failure function from the nearest neighbor representation of pattern $P$.

\begin{codebox}
\Procname{$\proc{KMP-Compute-Failure-Function2}(P, \nu_{p}(P), \nu_{n}(P))$}
\li $m \gets |P|$
\li $k \gets 0$
\li $\pi[1] \gets 0$
\li \For $q \gets 2$ \To $m$
\li \Do
        $(j_1,j_2) \gets (\nu_{p}(P)[k+1], \nu_{n}(P)[k+1])$
\li     \While $k > 0$ and ($P[q]<P[i-k+j_1-1]$ or $P[q]>P[i-k+j_2-1])$)
\li     \Do
            $k \gets \pi[k]$
\li         $(j_1,j_2) \gets (\nu_{p}(P)[k+1], \nu_{n}(P)[k+1])$
        \End
\li     $k \gets k+1$
\li     $\pi[q] \gets k$
    \End
\li \Return $\pi$
\end{codebox}

The time complexity of $\proc{KMP-Compute-Failure-Function2}$ is $O(m)$ from the linear scan of the pattern similarly to $\proc{KMP-Order-Matcher2}$.

\subsection{Correctness and Time Complexity}

The correctness of our optimized algorithm is derived from that of the previous $O(n \log m)$ algorithm since
the difference of the text search is only on rank comparison logic and each comparison result is the same as that of the previous one.
The same failure function $\pi$ is applied and the order-statistic tree $\mathcal{T}$ is only used to compute the nearest neighbor representation of $P$.



The time complexity of the overall algorithm is $O(n + m \log m)$: $O(m \log m)$ time for the computation of the nearest neighbor representation of the pattern, and $O(n)$ time
for text search, and $O(m)$ time for the construction of $\pi$ function.
$O(n + m \log m)$ is almost linear to the text length $n$ when $n$ is much larger than $m$, which is a typical case in pattern matching problems. The only non-linear factor $\log m$ comes from the representation of order relations.

\subsection{Generalized order-preserving matching}

A generalization of order-preserving matching is possible with some practical applications if we consider only the orders of the last $k$ characters \changeParkM{for a given $k \leq m$. For example, in the stock market scenario \changeParkO{in the Introduction} of finding a period when a share price of a company dropped consecutively for $10$ days and then went up for the next $5$ days, it is sufficient to compare each share price with \changeParkO{that} of the day before, which corresponds to $k=1$.}
\changeParkM{Our solution is easily applicable} to this generalized problem if the order-statistic tree $\mathcal{T}$ is maintained to keep only the last $k$ characters of the inserted characters. The time complexity of \changeParkM{the $O(n \log m)$ algorithm with prefix representation becomes $O(n \log k)$ and that of the $O(n + m \log m)$ algorithm with nearest neighbor representation $O(n + m \log k)$ since the number of characters in $\mathcal{T}$ is bounded to $k$.} Both time complexities are reduced to $O(n)$ if $k$ is a constant number.

\subsection{Remark on the alphabet size}

We have no restrictions on the numbers in $\Sigma$, insofar as a
comparison of two numbers can be done in constant time. In the case of
$\Sigma=\{1,2,\ldots,U\}$, however, the order-statistic tree in
\proc{Compute-Nearest-Neighbor-Rep} can be replaced by van Emde Boas tree~\cite{Pet75} or y-fast trie~\cite{Wil83} which takes $O(U)$ space and requires $O(\log\log U)$ time per operation.


\section{$O(n \log m)$ algorithm for multiple patterns}

In this section, we consider a generalization of order-preserving matching for multiple patterns.

\begin{definition}[Order-preserving matching for multiple patterns]
Given a text $T[1..n]$ and a set of patterns $\mathcal{P}=\{ P_1, P_2, ..., P_w \}$,
order-preserving matching for multiple patterns is the problem of finding all positions of $T$ matched with any pattern in $\mathcal{P}$.
\label{definition:order_preserving_matching_for_multiple_patterns}
\end{definition}

We propose a variant of the Aho-Corasick algorithm~\cite{Aho90} for the multiple pattern case whose time complexity is $O(n \log m)$ where $m$ is the sum of the lengths of the patterns.

\subsection{Prefix representation of Aho-Corasick automaton}
From the prefix representation of the given patterns, an Aho-Corasick automaton can be defined to match order relations.
The Aho-Corasick automaton consists of the following components.

\medskip
\begin{enumerate}
\item $Q$: a finite set of states where $q_0 \in Q$ is the initial state.
\item $g: Q \times \mathbb{N}_m \rightarrow Q \cup \{\text{fail}\}$: a forward transition function. $\mathbb{N}_m$ is the set of integers in $[1..m]$.
\item $\pi: Q \rightarrow Q$: a failure function.
\item $d: Q \rightarrow \mathbb{Z}$: the length of the prefix represented by each state $q$.
\item $P: Q \rightarrow \mathcal{P}$: a representative pattern of each state $q$ which has the prefix represented by $q$. If there are more than one such patterns, we use the pattern with the smallest index.
\item $out: Q \rightarrow \mathcal{P} \cup \{ \phi \}$: the output pattern of each state $q$. If $q$ does not match any pattern, $out[q]=\phi$, otherwise $out[q]=P_i$ for the longest pattern $P_i$ such that the prefix representation of $P_i$ is matched with that of any suffix of $P[q][1..d[q]]$.
\end{enumerate}

Given the set of patterns, an Aho-Corasick automaton of the prefix representations is constructed from a trie in which each node represents a prefix of the prefix representation of some pattern. The nodes of the trie are the states of the automaton and the root is the initial state $q_0$ representing the empty prefix.
Each node $q$ is an accepting state if $out[q] \neq \phi$, which means that $q$ corresponds to the prefix representation of the pattern $out[q]$.
The forward transition function $g$ is defined so that $g[q_i, \alpha] = q_j$ when $q_i$ corresponds to $\mu(P_k)[1..d[q_i]]$ and $q_j$ corresponds to $\mu(P_k)[1..d[q_i]+1]$ for some pattern $P_k$ where $\alpha = \mu(P_k)[d[q_i]]$. The trie can be constructed in $O(m)$ time once the prefix representation of the patterns are given.

\begin{figure}[t]
  \includegraphics[scale=0.55]{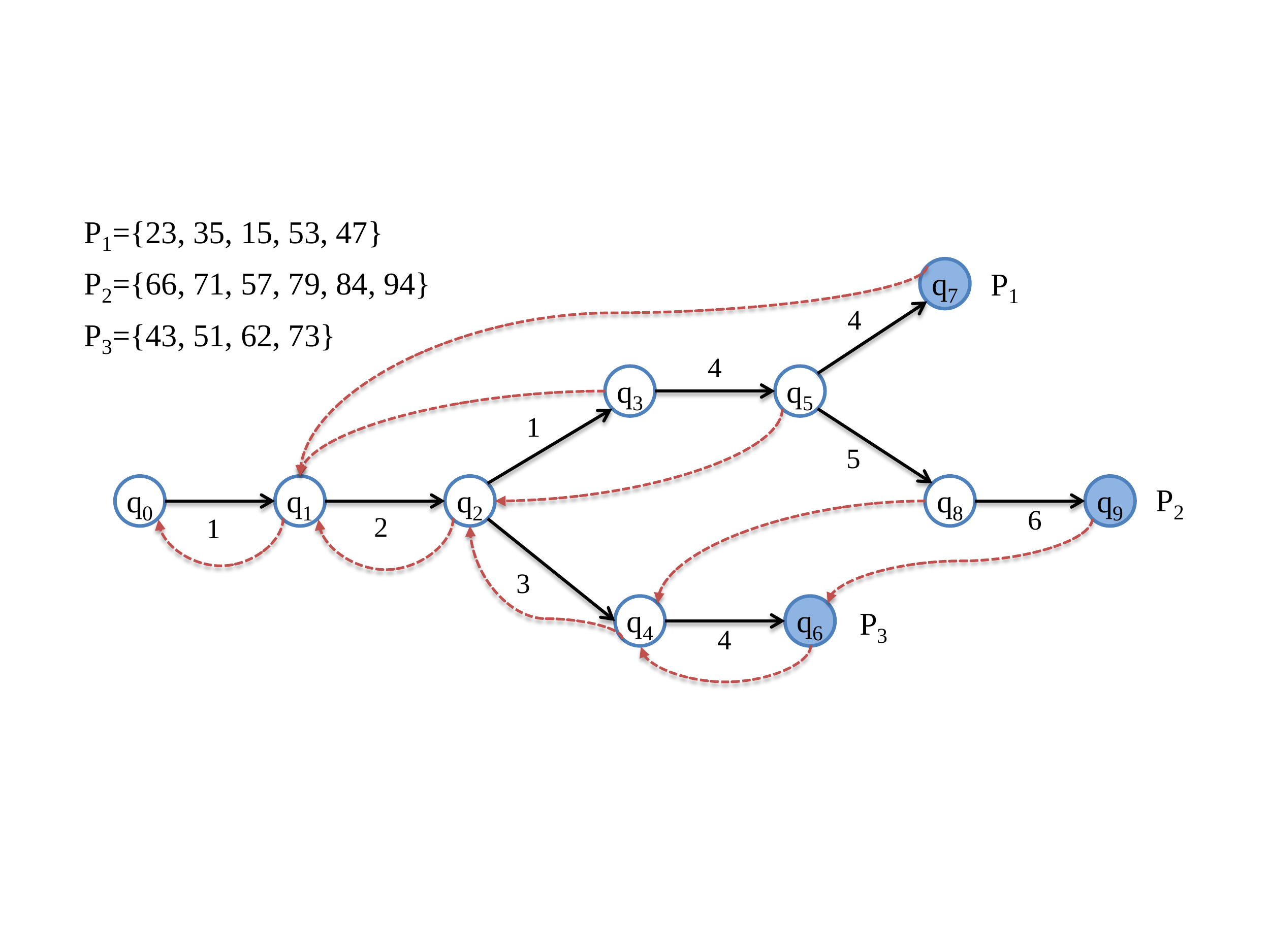}
  \caption{Example of AC automaton and failure function}
  \label{FIG:order_AC_automaton_resize}
\end{figure}

Fig.~\ref{FIG:order_AC_automaton_resize} shows an example of an Aho-Corasick automaton with three patterns $P_1=\{ 23, 35, 15, 53, 47 \}$, $P_2=\{ 66, 71, 57, 79, 84, 93 \}$, $P_3 = \{ 43, 51, 62, 73 \}$. The automaton is constructed from the prefix representations $\mu(P_1)=(1,2,1,4,4)$, $\mu(P_2)=(1,2,1,4,5,6)$ and $\mu(P_3)=(1,2,3,4)$ regardless of the pattern characters. For example, $q_5$ represents the prefix $(1,2,1,4)$ which matches with $\mu(P_1)$ and $\mu(P_2)$ even though $P_1[1..4]$ and $P_2[1..4]$ have different characters.

Compared to the original Aho-Corasick algorithm, we have two additional values $d[q]$ and $P[q]$ for each state $q$.
Both of them are recorded to maintain the order-statistic tree per pattern during the construction of the failure function $\pi$. The details are described in the following sections.

\subsection{Aho-Corasick failure function}

The failure function $\pi$ can be defined so that $\pi[q_i]=q_j$ if and only if the prefix represented by $q_j$ (i.e. $\mu(P[q_j])[1..d[q_j]]$) is the prefix representation of the longest proper suffix of $P[q_i]$ (i.e. $\mu(P[q_i][k..d[q_i]])$ for some $k$). For example, for $q_8$ in Fig.~\ref{FIG:order_AC_automaton_resize} with the prefix $(1,2,1,4,5)$ of $\mu(P_2)$, $\pi[q_8]=q_4$ because $P_2[3..5]$ is the longest proper suffix of $P_2$ whose prefix representation $(1,2,3)$ is the prefix of some pattern. Here, $P[q_4]=P_3$ and $\mu(P[q_4])[1..3]=(1,2,3)$ which is matched with $\mu(P_2[3..5])$.

\subsection{Text search}

A variant of the Aho-Corasick algorithm can be designed for the multiple pattern matching of order relations as in $\proc{AC-Order-Matcher-Multiple}$. Assuming that the prefix representations of all the patterns and the failure function are available, it scans the text and follows the Aho-Corasick automaton until there is no matched forward transition. Then, it follows the failure function until a successful forward transition is found.
In the initial state $q_0$, it \emph{never} fails to follow the forward transition because any character can be matched at the first character.
Whenever it reaches one of the accepting states, it outputs the position of the text and the matched pattern.

The order-statistic tree $\mathcal{T}$ is maintained to compute each rank value adaptively. For every forward transition, $T[i]$ is inserted to $\mathcal{T}$, and for every backward transition $\pi[q_i]=q_j$, the oldest $d[q_i]-d[q_j]$ characters are deleted from $\mathcal{T}$. The rank of $T[i]$ should be calculated again for each backward transition after $\mathcal{T}$ is properly updated. For example, when $\proc{AC-Order-Matcher-Multiple}$ reaches state $q_3$ of Fig.~\ref{FIG:order_AC_automaton_resize} after reading the first three characters from the text $(20, 30, 10, 15)$, $\mathcal{T}$ contains $\{ 20, 30, 10 \}$ that is the prefix of the text represented by $q_3$. As there is no forward transition from $q_3$ that matches the rank $2$ of the next character $15$, the state is changed to $q_1$ by following the failure transition. The oldest $d[q_3]-d[q_1]=2$ characters are deleted from $\mathcal{T}$ so that it contains $\{ 10 \}$ at the next step. The state is then changed to $q_2$ by following the forward transition $2$ with inserting $15$ to $\mathcal{T}$ (which is rank 2 in $\{ 10, 15 \}$).

\begin{codebox}
\Procname{$\proc{AC-Order-Matcher-Multiple}(T, \mathcal{P})$}
\li $n \gets |T|$
\li \For $i \gets 1$ \To $w$
\li \Do
        $\mu(P_i) \gets \proc{Compute-Prefix-Rep}(P_i)$
    \End
\li $(\pi, out) \gets \proc{Compute-AC-Failure-Function}(\mathcal{P})$
\li $\mathcal{T} \gets \phi$
\li $q \gets q_0$
\li \For $i \gets 1$ \To $n$
\li \Do
        $\proc{OS-Insert}(\mathcal{T}, T, i)$
\li     $r \gets \proc{OS-Rank}(\mathcal{T}, T[i]$)
\li     \While $g[q, r] \isequal fail$
                                            \label{li:AC-Order-Matcher-Multiple-while}
\li     \Do
            $\proc{OS-Delete}(\mathcal{T}, T[i-d[q]..i-d[\pi[q]]-1])$
\li         $q \gets \pi[q]$
\li         $r \gets \proc{OS-Rank}(\mathcal{T}, T[i])$
        \End
\li     $q \gets g[q, r]$
\li     \If $out[q] \neq \phi$
\li     \Then
            print ``pattern'' $out[q]$ ``occurs at position'' $i$
        \End
    \End
\end{codebox}

The time complexity of $\proc{AC-Order-Matcher-Multiple}$ is $O(n \log m)$ except the preprocessing of the patterns because it does $n$ insertions in $\mathcal{T}$ and thus at most $n$ deletions can take place. Checking $g[q, r]$ in line~\ref{li:AC-Order-Matcher-Multiple-while} takes $O(\log m)$ time as well. As each operation takes $O(\log m)$ time and there are $O(n)$ operations, the total time is $O(n \log m)$.

\subsection{Construction of Aho-Corasick failure function}

$\proc{Compute-AC-Failure-Function}$ shows the construction algorithm of the Aho-Corasick failure function.
As in the original Aho-Corasick algorithm, it computes the failure function in the breadth first order of the automaton.

The main difference from the original Aho-Corasick algorithm is that we maintain multiple order-statistic trees simultaneously (one per pattern) because the rank value of a character depends on the pattern in which the rank is calculated. Let $\mathcal{T}(P_i)$ denote the order-statistic tree for the pattern $P_i$, and let's assume that a representative pattern $P[q]$ is recorded for each node $q$ such that $q$ is reachable by some prefix of the prefix representation of $P[q]$.

We maintain each order-statistic tree $\mathcal{T}(P[q])$ of $P[q]$ so that it contains the characters of the longest proper suffix of $P[q][1..d[q]]$ whose prefix representation is a prefix of the prefix representation of some pattern. Let's consider a forward transition $g[q_i, \alpha]=q_j$ such that $\pi[q_i]$ is available but $\pi[q_j]$ is to be computed. If $P[q_i]=P[q_j]$, $\mathcal{T}(P[q_i])=\mathcal{T}(P[q_j])$ and $\mathcal{T}(P[q_j])$ already contains the characters of $P[q_j]$. It can be updated by inserting $P[q_j][d[q_j]]$ and deleting some characters from $\mathcal{T}(P[q_j])$. However, if $P[q_i] \neq P[q_j]$, we should initialize $\mathcal{T}(P[q_j])$ by inserting characters
of the suffix of $P[q_j][1..d[q_j]-1]$ so that it has the same number of characters as $\mathcal{T}(P[q_i])$. $\mathcal{T}(P[q_j])$ then can be updated as in the other case. In both cases, the rank of $P[q_j][d[q_j]]$ in $\mathcal{T}(P[q_j])$ is computed again to find the correct forward transition starting from $\pi[q_i]$.


For instance, let's consider node $q_5$ in Fig.~\ref{FIG:order_AC_automaton_resize}. $P[q_5]=P_1$ and $\mathcal{T}(P_1)$ has $\{ 15, 53 \}$ since $d[\pi[q_5]]=2$.
When $\pi[q_7]$ is computed, it inserts $47$ to $\mathcal{T}(P_1)$ which is rank $2$ in $\{ 15, 53, 47 \}$ and tries to follow the rank $2$ from $\pi[q_5]=q_2$. As there is no forward transition of $q_2$ with label $2$, it follows the failure function $\pi[q_2]=q_1$ and deletes $15$ from $\mathcal{T}(P_1)$. Similarly, there is no forward transition of the rank $1$ of $47$ in $\{ 53, 47 \}$ from $q_1$, it reaches $q_0$. Finally, it follows the forward transition of $q_1$ by the rank $1$ of $47$ in $\{ 47 \}$ and $\pi[q_7]=q_1$. On the other hand, when $\pi[q_8]$ is computed, $P[q_8]=P_2$ and $P[q_8] \neq P[q_7]$. The last $d[\pi[q_5]]$ characters of $P_2[1..d[q_5]]$ are inserted to $\mathcal{T}(P_2)$, and $\mathcal{T}(P_2)$ becomes $\{ 57, 79 \}$. Then, the next character $84$ of $P[q_8]$ is inserted to $\mathcal{T}(P_2)$ that is rank $3$ of $\{ 57, 79, 84 \}$ and it follows the rank $3$ from $q_2$, which results in $\pi[q_8]=q_4$.


\begin{codebox}
\Procname{$\proc{Compute-AC-Failure-Function}(T, \mathcal{P})$}
\li $\pi[q_0] \gets q_0$
\li \For \textbf{each} $P_i \in \mathcal{P}$
\li \Do
        $\mathcal{T}(P_i) \gets \phi$
\li     $out[q_i] \gets P_i$ for the last state $q_i$ of $P_i$
    \End
\li \For \textbf{each} $q_i \in Q$ (BFS order)
\li \Do
        \For \textbf{each} $\alpha$ such that $g[q_i, \alpha] \neq \text{fail}$
\li     \Do
            $q_j \gets g[q_i, \alpha]$, $c \gets P[q_j][d[q_j]]$
                                                                        \label{li:Compute-AC-Failure-Function-forward1}
\li         \If $P[q_i] \neq P[q_j]$
\li         \Do
                \For $k \gets 1$ \To $d[\pi[q_i]]$
\li             \Do
                    $\proc{OS-Insert}(\mathcal{T}(P[q_j]), P[q_j], d[q_i]-d[\pi[q_i]]+k)$     \label{li:Compute-AC-Failure-Function-insert1}
                \End
            \End
\li         $\proc{OS-Insert}(\mathcal{T}(P[q_j]), P[q_j], d[q_j])$                                         \label{li:Compute-AC-Failure-Function-insert2}
\li         $r \gets \proc{OS-Rank}(\mathcal{T}(P[q_j]), c$)
\li         $q_p \gets q_i$, $q_h \gets \pi[q_i]$
\li         \While $g[q_h, r] \isequal fail$
                                                                        \label{li:Compute-AC-Failure-Function-forward2}
\li         \Do
                $\proc{OS-Delete}(\mathcal{T}(P[q_j]), P[q_j][i-d[q_p]+1..i-d[q_h]])$
\li             $r \gets \proc{OS-Rank}(\mathcal{T}(P[q_j]), c$)
\li             $q_p \gets q_h$, $q_h \gets \pi[q_h]$
            \End
\li         $\pi[q_j] \gets g[q_h, r]$
\li         \If $out[q_j] \isequal \phi$
\li         \Do
                $out[q_j] \gets out[\pi[q_j]]$
            \End
        \End
    \End
\li \Return $(\pi, out)$
\end{codebox}


The time complexity of $\proc{Compute-AC-Failure-Function}$ can be analyzed as follows. The number of all the forward transitions is at most $m$ and there are at most $m$ insert operations on $\mathcal{T}$ because each character of a pattern can be inserted either in line~\ref{li:Compute-AC-Failure-Function-insert1} or in line~\ref{li:Compute-AC-Failure-Function-insert2} but cannot be in both.
The number of deleted characters cannot exceed the number of inserted characters and the number of rank computations is also bounded by $m$.
As the number of each operation is $O(m)$ and each takes $O(\log m)$, the total time complexity is $O(m \log m)$.

\subsection{Correctness and Time Complexity}

The correctness of our algorithm can be easily derived from the correctness of the original Aho-Corasick algorithm and our version for the single pattern matching.

The total time complexity is $O(n \log m)$ due to $O(m \log m)$ for prefix representation and failure function computation, $O(n \log m)$ for text search.
Compared with $O(n \log |\Sigma|)$ time of the exact pattern matching where $\Sigma$ is the alphabet, our algorithm has a comparable time complexity
since $|\Sigma|$ for numeric strings can be as large as $m$.

Note that we cannot remove $\log m$ factor from the above time complexity as in the single pattern case since
$O(\log m)$ time has to be spent at each state to find the forward transition to follow even with the nearest neighbor representation.

%

\section{Conclusion}

We have introduced \emph{order-preserving matching} and defined
\emph{prefix representation} and \emph{nearest neighbor representation} of order relations of a numeric string. By using these representations, we developed an $O(n + m \log m)$ algorithm for single pattern matching and an $O(n \log m)$ algorithm for multiple pattern matching. We believe that our work opens a new direction in string matching of numeric strings with many practical applications.





\bibliographystyle{abbrv}
\bibliography{order_preserving_matching}







\end{document}